\documentclass[preprint,aps,prc,showpacs,nofootinbib,secnumarabic,floatfix]{revtex4-1}

\usepackage{graphicx}
\usepackage{bm}
\usepackage{color}
\usepackage{gensymb}
\usepackage{amsmath}
\usepackage{slashed}
\usepackage{appendix}
\usepackage{hyperref}

\newcommand{\ltsim}{\protect\raisebox{-0.5ex}{$\:\stackrel{\textstyle <}
	{\sim}\:$}}

\newcommand{\bvec}[1]{\ensuremath{\boldsymbol{#1}}}

\begin{document}

\title{Translationally invariant shell model calculation of the quasielastic $(p,2p)$ process
       at intermediate relativistic energies}

\author{A.B. Larionov$^1$\footnote{e-mail: larionov@theor.jinr.ru},
        Yu.N. Uzikov$^{2,3,4}$\footnote{e-mail: uzikov@jinr.ru}}
  
\affiliation{$^1$ Bogoliubov Laboratory of Theoretical Physics, Joint Institute for Nuclear Research, 141980 Dubna, Russia\\
             $^2$ Laboratory of Nuclear Problems, Joint Institute for Nuclear Research, 141980 Dubna, Russia\\
             $^3$ Department of Physics, Moscow State University, 119991 Moscow, Russia\\
             $^4$ Dubna State University, 141980 Dubna, Russia}

\begin{abstract}
  Relativistic beams of heavy ions interacting with various nuclear targets allow to study a broad range of problems
  starting from nuclear equation of state to the traditional nuclear structure. Some questions which were impossible
  to answer heretofore -- can be addressed nowadays by using inverse kinematics. These includes the structure of short-lived nuclei
  and the precision study of exclusive channels with production of residual nuclei in certain quantum states. 
  Theoretical understanding such processes is so far based on factorization models which combine the single-step amplitude
  of the reaction on a bound nucleon or nuclear cluster with a certain wave function of its relative  motion  with respect
  to the residual nucleus. The nuclear structure information is encoded in the spectroscopic amplitude, calculable within nuclear many-body
  theories. In this work, we use for this purpose the translationally-invariant shell model with configuration mixing
  and demonstrate that it successfully reproduces the single-differential and integrated cross sections of the quasielastic proton knockout,
  $^{12}\mbox{C}(p,2p)^{11}\mbox{B}$, with outgoing $^{11}$B in the ground state and low-lying excited states measured at GSI
  at 400 MeV/nucleon.  
\end{abstract}

\maketitle

\section{Introduction}
\label{intro}

Quasielastic (QE) knock-out reactions are the most direct way to access the momentum distribution
of the valence nucleons given  by the square of their wave function  (WF) in momentum space.
In the low-momentum region, the WFs are determined by nuclear mean field potential.
Distortions of the incoming and outgoing proton waves including absorption effects
are governed by nuclear optical potential that is mostly imaginary at high momenta and proportional to the local nuclear density.    
The nuclear mean field and optical potentials are rather well known for ordinary stable nuclei but represent a major uncertainty for exotic ones
close to the neutron drip line. The studies of the structure of exotic nuclei using inverse kinematics with proton target at rest
are in a focus of experiments at RIKEN \cite{Kondo:2009zza} and GSI \cite{R3B:2019gix}. One of the most important questions is the quenching of single-particle
strength and the dependence of this effect on the isospin asymmetry, see Ref.~\cite{Aumann:2020tcq} for a recent review.
As a first step, before being extended towards exotic nuclear region, any theoretical model of $(p,pN)$ reactions should be tested for $\beta$-stable nuclear beams
where a number of  uncertainties in the model parameters is minimal.
 
In this work, we address the proton knock-out reaction $^{12}\mbox{C}(p,2p)^{11}\mbox{B}$ measured in Ref.~\cite{Panin:2016div} with 400 MeV/nucleon $^{12}$C beam
colliding with proton target. We apply the translationally-invariant shell model (TISM) \cite{NS} which allows to calculate the spectroscopic amplitudes
of the virtual decay $^{12}\mbox{C} \to p\, ^{11}\mbox{B}$ for the given relative WF of the proton and residual nucleus and internal state of the residual nucleus.
The present study is complementary to our previous work \cite{Larionov:2023cpe} where the TISM has been used to analyze the proton knock-out from a short-range correlated
$pN$ pair in the $^{12}$C nucleus by a proton that yielded a good agreement with the BM@N data \cite{Patsyuk:2021fju}.

In sec.~\ref{TISM} we give a brief description of the TISM in the harmonic oscillator (HO) basis, with an emphasis on how it differs from the conventional shell model.
Configuration mixing is accounted for within the intermediate coupling model \cite{Balashov:1964,Boyarkina73}.
In sec.~\ref{model}, the basic elements of the reaction
model are described starting from the amplitude in the impulse approximation (IA) and
then adding the initial- and final state interactions (ISI/FSI) in the eikonal approximation.  Sec.~\ref{results} contains results of our calculations of $^{12}\mbox{C}(p,2p)^{11}\mbox{B}$ process
at 400 MeV/nucleon in comparison with experimental data \cite{Panin:2016div}. In sec.~\ref{discuss}, we discuss various calculations of the
spectroscopic factor and other sources of theoretical uncertainties. Sec.~\ref{summary} summarizes the main results of the present work.  
Appendix \ref{FPC} contains the derivation of the relation between the fractional parentage coefficient of the TISM and that of the conventional shell model.   

\section{The translationally-invariant shell model}
\label{TISM}

The most complicated and important element of any reaction model is the spectroscopic amplitude of a virtual  transition,
\begin{equation}
  A \to B + X~,   \label{A2BX}
\end{equation}
of the nucleus $A$ into residual nucleus $B$ and cluster $X$ both being in definite internal states and a certain state of relative motion.
The TISM, which has been developed starting from 60's in Moscow State University
\cite{NS,Balashov:1964,Kurdyumov:1970qkz, Boyarkina73,Smirnov:1977zz}, provides an efficient and computationally economic way to evaluate the spectroscopic amplitude
in the case of $1p$-shell nuclei.
Although the TISM formalism is quite old and well known, we will briefly describe it here.

In the ordinary shell model, the center of mass (c.m.) of the nucleus
is not at rest and is not in a state of uniform rectilinear motion, but oscillates relative to an arbitrary point in space
chosen as the origin of coordinates.
The set of excited states of the shell model includes the so called "spurious" states in which all the energy is concentrated in the excitation of the nuclear
c.m. oscillations without changing the internal motion of the nucleus.
These spurious excitations of the c.m. of the nucleus are a serious drawback of the ordinary nuclear shell model
and should be excluded when performing calculations of physical observables, most importantly, for light nuclei.

A time-honored way to remove spurious excitations is to use the HO potential well that allows to separate
the Hamiltonian of the particle system into the Hamiltonian of the c.m. motion and
the Hamiltonian of internal motion.
The latter is used to construct the TISM.
Thus, this model operates with internal WFs which describe harmonic oscillations of nucleons with respect to the position of their c.m..  
Equivalently, the internal WFs can be considered as the solutions of a many-body Schr\"odinger equation for the system of
particles interacting through the two-body HO potentials
(see Ref.~\cite{Kurdyumov:1970qkz} and references therein).
The complete basis set of the TISM WFs can be used to take into account different phenomenological interactions
(see discussion in the end of this section).

According to Refs.~\cite{Kretzschmar60,Kurdyumov:1970qkz,ED}, in the TISM
for a nucleus with $A$ nucleons, the classification of the orbital WFs 
is based on the chain of groups $U_{3(A-1)}=U_1 \times SU_{3(A-1)}$,
$SU_{3(A-1)} \supset SU_3 \times SU_{A-1} \supset R_3 \times S_A$.
Thus, each state of the nucleus is characterized by the total number of HO
quanta $N$, the Young's scheme $[f]$, the Elliott symbol $(\lambda\mu)$,
and the total orbital angular momentum $L$
(for the irreducible representations of the groups $U_1$, $S_A$, $SU_3$, and $R_3$, respectively).
In addition, when one constructs the total coordinate-spin-isospin WF, 
the spin $S$ and isospin $T$ quantum numbers are introduced  (for the irreducible representations of the corresponding $SU_2$ groups which appear in the chain
$SU_{4A} \supset SU_4 \times SU_A \supset SU_2 \times SU_2 \times S_A$).  
The total antisymmetric WF is denoted by the symbol $|A N [f]\kappa (\lambda\mu) LST M_L M_S M_T> $,
where $M_L, M_S, M_T$ are the projections of $L$, $S$ and $T$, respectively, and $\kappa$ represents additional
(if any) quantum numbers necessary for the unambiguous classification of the states.
This function depends on a certain set of $3(A-1)$ Jacobi coordinates
and does not depend on the coordinate of the c.m. of the nucleus.
The TISM WF is given  by the following sum of  the products
of fully antisymmetric internal WFs of the nucleus $B$, which contains $A-b$ nucleons with
numbers $1,2,\dots, A-b$, and  the cluster $X$ containing $b$ nucleons  with numbers
  $A-b+1,A-b+2, \dots, A$  multiplied by  their relative WF $\psi_{n\Lambda}^{M_\Lambda}$
(see Eqs.(VII.14),(VII.17) in Ref.~\cite{NS} or Eqs.(19),(21) in
Ref.~\cite{Kurdyumov:1970qkz} for $b=1,2$, or Eq.(3) in Ref.~\cite{Kudeyarov:1971gdd}
for arbitary $b$):
\begin{eqnarray}
&&  |A \alpha LST M_L M_S M_T  \rangle =
\sum (L_B M_{L_B} {\cal L} M_{\cal L} | L M_L)\, (\Lambda M_\Lambda L_X M_{L_X} | {\cal L} M_{\cal L}) \nonumber \\
    && \times (S_B M_{S_B} S_X M_{S_X} | S  M_S)\, (T_B M_{T_B} T_X M_{T_X} | T M_T)  \nonumber \\
    && \times
  \langle A \alpha LST | (A-b) \alpha_B L_BS_BT_B; n\Lambda, b \alpha_X  L_XS_XT_X\{{\cal L}\} \rangle \nonumber \\
    && \times
   |(A-b) \alpha_B L_BS_BT_B M_{L_B} M_{S_B} M_{T_B} \rangle \nonumber \\
   && \times \psi_{n\Lambda}^{M_\Lambda}(\bvec{R}_B - \bvec{R}_X)\,
   |b \alpha_X L_XS_XT_X M_{L_X} M_{S_X} M_{T_X} \rangle~, \label{defFPC}
\end{eqnarray}
where  we used the standard notations for the Clebsch-Gordan coefficients for the $R_3$-group.
For brevity, following Ref.~\cite{Kurdyumov:1970qkz},
we included the label $\alpha \equiv N[f] \kappa (\lambda\mu)$.
In Eq.(\ref{defFPC}), the WF $\psi_{n\Lambda}^{M_\Lambda}(\bvec{R}_B - \bvec{R}_X)$
depends on the position vectors of the c.m. of the nucleus $B$ ($\bvec{R}_B$) and the cluster $X$ ($\bvec{R}_X)$
and, thus, describes the state of their relative motion with number of HO quanta $n$, the relative orbital momentum
$\Lambda$ and the magnetic quantum number, $M_\Lambda$. 
The factor
$\langle A \alpha LST | (A-b) \alpha_B L_BS_BT_B; n\Lambda, b \alpha_X  L_XS_XT_X\{{\cal L}\} \rangle$
is the fractional parentage coefficient (FPC) of the TISM, while the sum runs over all quantum numbers (except those
that appear in the internal WF of the nucleus $A$) under the necessary selection rules.
\footnote{This implies, in particular, the summation over $N_B, n, N_X$ under the condition $N=N_B+n+N_X$.}

The  WF of the nucleus $A$ with definite total angular momentum $J$ and its $z$-projection $M$  is 
constructed by using vector coupling $\bvec{J}=\bvec{L}+\bvec{S}$:
\begin{equation}
\label{WFAJ}
  \Psi_A =  |A \alpha LS(J)T M M_T \rangle \equiv \sum_{M_L,M_S} |A \alpha LST M_L M_S M_T \rangle
      (L M_L S M_S | J M)~.
\end{equation}
The WFs for the nucleus $B$, $\Psi_B$,  and cluster $X$,   $\Psi_X$,
with the total angular momenta and their $z$-projections
$J_BM_B$, $J_XM_X$, respectively, have  similar forms
constructed by using vector couplings $\bvec{J}_B=\bvec{L}_B+\bvec{S}_B$, $\bvec{J}_X=\bvec{L}_X+\bvec{S}_X$.
For the WF (\ref{WFAJ}) the fractional parentage expansion in the products of $\Psi_B$
and $\Psi_X$, can be written similarly to Eq.(\ref{defFPC}) with the same FPCs.

The spectroscopic amplitude  $S_A^X$ for the transition from the state  (\ref{WFAJ})
to the state  $\Psi_B$ of the residual nucleus $B$ with the emission of the cluster $X$
in the state  $\Psi_X$ and definite state of their relative motion $\Psi_{n\Lambda}^{M_\Lambda}(\bvec{R}_B - \bvec{R}_X)$
is defined as the following overlap integral \cite{Kurath:1974afb,Smirnov:1977zz}:
\begin{equation}
    S_A^X
         = \left(\begin{array}{c}
    A \\
    b
         \end{array}\right)^{1/2}
       \langle \Psi_B, \Psi_{n\Lambda}^{M_\Lambda}(\bvec{R}_B - \bvec{R}_X), \Psi_X | \Psi_A \rangle~,   \label{S_A^X_def}
 \end{equation}
where the first  factor takes into account the identity of the nucleons.
Using Eqs.(\ref{defFPC}),(\ref{WFAJ}), the spectroscopic amplitude can be written as follows
(see Refs.~\cite{Smirnov:1977zz,Uzikov:2022epe}):
 \begin{eqnarray}
   S_A^X     &=&
        \left(\begin{array}{c}
    A \\
    b
       \end{array}\right)^{1/2}
       \sum_{{\cal L} J_0 M_0}
   \left\{\begin{array}{lll}
              L_B      & S_B & J_B \\
              {\cal L} & S_X & J_0 \\
               L       & S   & J
          \end{array}
   \right\} \sqrt{(2L+1)(2S+1)(2J_B+1)(2J_0+1)}\, \nonumber \\
  && \times \langle A \alpha LST|(A-b) \alpha_B  L_BS_BT_B; n\Lambda, b \alpha_X L_XS_XT_X\{{\cal L}\} \rangle
   U(\Lambda L_X J_0 S_X;{\cal L} J_X)
  \nonumber \\
  && \times  (J_B M_B J_0 M_0 | J M)\, (\Lambda M_\Lambda J_X M_X | J_0 M_0)\,
            (T_B M_{T_B} T_X M_{T_X} | T M_T)~.      \label{S_A^X}
\end{eqnarray}
Here we introduced the $9j$-symbols and Racah coefficients in the standard notations.
This equation gives the solution for the problem of finding the virtual decay amplitude (\ref{A2BX}).

The FPCs of TISM which enter Eq.(\ref{S_A^X}) can be related to the FPCs of the conventional HO shell
model \cite{Kurdyumov:1970qkz,Smirnov:1977zz}.
For the case when the states of the nuclei $A$ and $B$ contain the minimum numbers of the
oscillator quanta $N=A-4$ and $N_B=A-b-4$ the relation takes the following form (see Eq.(4) in Ref.~\cite{Smirnov:1977zz}):
\begin{eqnarray}
  && \langle A \alpha LST | (A-b) \alpha_B L_BS_BT_B; n\Lambda, b \alpha_X L_XS_XT_X\{{\cal L}\} \rangle     \nonumber \\
  && = (-1)^n \left(\frac{A}{A-b}\right)^{n/2}
  \left(\begin{array}{c}
    A-4 \\
    b
  \end{array}\right)^{1/2}
  \left(\begin{array}{c}
    A \\
    b
  \end{array}\right)^{-1/2}  \nonumber \\
  && \times \langle p^{A-4} \alpha LST | p^{A-b-4} \alpha_B L_BS_BT_B; p^b \alpha_X {\cal L}S_XT_X \rangle \nonumber \\
  && \times  \langle p^b \alpha_X {\cal L}S_XT_X | n\Lambda,  b \alpha_X L_XS_XT_X \rangle~,  \label{FPC_TISM_vs_usual}
\end{eqnarray}
where the last factor is called a ``cluster coefficient'' which is the overlap integral of the WF of $b$ $p$-shell nucleons with
the WF $\psi_{n\Lambda}(\bvec{R}_X)$ times the internal WF of the cluster $X$, $|b \alpha_X L_XS_XT_X \rangle$.
It is assumed that the last two WFs are vector-coupled according to $\bvec{\cal L} =  \bvec{\Lambda} + \bvec{L}_X$.
The derivation of this relation was not present  explicitly  in Ref.~\cite{Smirnov:1977zz} and  is
given in the Appendix \ref{FPC}.
For $p$-shell nuclei, the FPCs of the conventional HO shell model for $b=1,2$ can be calculated by standard techniques \cite{Jahn51,Elliott53}.

In the HO shell model, the nuclear states are highly degenerate which is far from the actual spectra of nuclear levels. 
It is well known that the residual nuclear interactions need to be taken into account, thereby resulting in the decomposition
of the actual many-body WF in the basis of the shell model WFs. This can be, in principle, done either with $LS$ \cite{PhysRev.51.95,PhysRev.51.597}, $jj$ \cite{PhysRev.88.804}
or intermediate \cite{Kurath:1956zz} coupling schemes which differ with respect to the assumed strength of the spin-orbit interaction.  

The model with intermediate coupling \cite{Kurath:1956zz,PhysRev.106.975,Cohen:1965qa,Cohen:1967zzb,Boyarkina73} turned out to be the most realistic phenomenological model
in the mass range $A=5-16$ that gives a reasonably good description of the nuclear energy levels, magnetic dipole moments, probabilities for
$M1$ gamma transitions and $\beta$-decay. We will apply here the version of the intermediate coupling model of Ref.~\cite{Boyarkina73}.
In Ref.~\cite{Boyarkina73}, the nuclear Hamiltonian is written as
\begin{equation}
   \hat{H} = \sum_{i=1}^A \hat{H}_i + \sum_{i < j} \hat{V}_{ij} + a \sum_{i=1}^A \hat{\bvec{l}}_i \hat{\bvec{s}}_i~, \label{H}
\end{equation}
where the first term represents the total single-particle energy of the nucleons in the pure HO shell model.
The second term gives the energy of
the pair interactions between nucleons of the $1p$-state. The two-body potential is expressed as follows:
\begin{equation}
  \hat{V}_{12} = [W + M \hat{P}_x + B \hat{P}_{\sigma} + H \hat{P}_x\hat{P}_{\sigma}] V(r_{12})~,    \label{V_12}
\end{equation}
where $\hat{P}_x$ and $\hat{P}_{\sigma}$ are, respectively, the space and spin exchange operators of the particles 1 and 2.
The Wigner, Majorana, Bartlett, and Heisenberg interaction constants are $W=-0.13$, $M=0.93$, $B=0.46$, and $H=-0.26$
corresponding to a so called 'Rosenfeld variant' (see Eq.(1) in Ref.~\cite{RevModPhys.25.390}).
The last term in the r.h.s. of Eq.(\ref{H}) is the spin-orbit potential.

The functional form of $V(r_{12})$ in Eq.(\ref{V_12}) is not explicitly given in Ref.~\cite{Boyarkina73}, since it is well known that 
the matrix elements of the central two-body interaction of the type $\langle  p^2 [f_X]^{2T_X+1,2S_X+1}{\cal L}| \hat{V}_{12} | p^2 [f_X]^{2T_X+1,2S_X+1}{\cal L} \rangle$
can be expressed via radial integrals $L$ and $K$ (see Eq.(9) in Ref.~\cite{PhysRev.51.95}):
\begin{eqnarray}
  L &\equiv& \left(\frac{3}{4\pi}\right)^2 \int d^3r_1 d^3r_2 \frac{x_1^2}{r_1^2} R_p^2(r_1)  V(|\bvec{r}_1-\bvec{r}_2|) \frac{x_2^2}{r_2^2} R_p^2(r_2)~, \label{L} \\
  K &\equiv& \left(\frac{3}{4\pi}\right)^2 \int d^3r_1 d^3r_2 \frac{x_1y_1}{r_1^2} R_p^2(r_1) V(|\bvec{r}_1-\bvec{r}_2|) \frac{x_2y_2}{r_2^2} R_p^2(r_2)~, \label{K}
\end{eqnarray}
where
\begin{equation}
  R_p(r) = \left(\frac{8}{3\pi^{1/2}r_0^5}\right)^{1/2} r\, \mbox{e}^{-r^2/2r_0^2} \label{R_p}
\end{equation}
is the radial WF of the $1p$-state nucleon in the ordinary HO shell model.
(For example, $\langle p^2 [2]^{2T_X+1,2S_X+1}D | V(|\bvec{r}_1-\bvec{r}_2|) | p^2 [2]^{2T_X+1,2S_X+1}D \rangle = L-K$.)

If one applies the parameterization
\begin{equation}
  V(r_{12}) = V_0 \exp[-(r_{12}/d)^2]~,     \label{Vr}
\end{equation}
then the following relations hold (see Eqs.(2) in Ref.~\cite{PhysRev.88.804}):
\begin{eqnarray}
   K &=&  V_0 (r_0/d)^4 [1 + 2(r_0/d)^2]^{-7/2}~,    \label{K} \\
   L/K &=& 3 + (d/r_0)^4 [1 + 2(r_0/d)^2]~.        \label{L/K}
\end{eqnarray}
The value $L/K$ can be considered as a measure of the ratio $\rho$ of the nuclear radius, $\sqrt{2}r_0$,
to the interaction radius, $d$, \cite{Kurath:1956zz}.
In the limit when the interaction radius is large, one has $L=V_0$ and $K=0$.
The nuclear levels of not very high excitation energy ($\ltsim 4$ MeV) with the same isotopic spin as the ground state
are only weakly sensitive to $L/K$ in the range $5-8$, corresponding to $\rho \simeq 1.1-1.6$, see Ref.~\cite{Kurath:1956zz}.
In Ref.~\cite{Boyarkina73}, the value $L/K=6$ was fixed while the values of the spin-orbit parameter $a$ and radial integral $K$ were chosen to reproduce 
the low-lying states in nuclei in the best way and, thus, 
vary from nucleus to nucleus. For $A=12$ and $A=11$ considered here, the values $a=-5$ MeV and $K=-1.2$ MeV
were chosen.
By using Eqs.(\ref{K}),(\ref{L/K}) one can then reconstruct values $d=r_0$ and $V_0=-56.12$ MeV. 
In calculations we use the parameter of the ordinary HO shell model $r_0=1.581$ fm \cite{Alkhazov:1972oie}.

The nuclear energy levels and the corresponding WFs were calculated in Ref.~\cite{Boyarkina73}
by diagonalizing the Hamiltonian (\ref{H}) on the basis of TISM WFs.
The resulting internal WFs are the linear superpositions of the TISM WFs:
\begin{equation}
\label{GTISM}
   \Psi_A^{J,T} =\sum_i \alpha_{[f_i]L_iS_i}^{A,JT}\,|A [f_i] L_iS_i(J)T \rangle~.
\end{equation}
For the nuclei $6\leq A\leq 14$ the coefficients $\alpha_{[f]LS}^{A,JT}$ have been tabulated in Ref.~\cite{Boyarkina73}.
They are real-valued and satisfy the normalization condition:
\begin{equation}
  \sum_i (\alpha_{[f_i]L_iS_i}^{A,JT})^2 =1~.    \label{norm_alpha}
\end{equation}
For the purposes of the present work, Table~\ref{tab:12C} lists the coefficients $\alpha$  for $^{12}$C ground state
and Table~\ref{tab:11B} -- for $^{11}$B ground state and two excited states.

\begin{table}[htb]
  \caption{\label{tab:12C} Contributing $(1p)^8$ TISM states denoted as $[f]^{(2T+1) (2S+1)}L$ with corresponding coefficients $\alpha_{[f]LS}^{A,JT}$
    for the $^{12}$C ground state ($J=T=0$). Taken from Ref.~\cite{Boyarkina73}.}
    \begin{center}
    \begin{tabular}{lllll}
    \hline
    \hline
    $[44]^{11}S$~~~ & $[431]^{13}P$~~~ & $[422]^{11}S$~~~ & $[422]^{15}D$~~~ & $[332]^{13}P$ \\
    0.840          & 0.492           & 0.064           & -0.200          & 0.086 \\
    \hline
    \hline
    \end{tabular}
  \end{center} 
\end{table}

\begin{table}[htb]
  \caption{\label{tab:11B} Same as in Table~\ref{tab:12C} but for $(1p)^7$ TISM states in $^{11}$B. The ground state and two excited states are included.
    The experimental and theoretical (in parentheses) values of excitation energy are taken,
    respectively, from Ref.~\cite{Ajzenberg-Selove:1990fsm} and Ref.~\cite{Boyarkina73}.}
    \begin{center}
              $E^*=0$ MeV, $T=1/2, J=3/2$ 
    \begin{tabular}{lllllll}
    \hline
    \hline
    $[43]^{22}P$~ & $[43]^{22}D$~ & $[421]^{22}P$~ & $[421]^{24}P$~ & $[421]^{22}D$~ & $[421]^{24}D$~ & $[421]^{24}F$ \\
    0.636        & 0.566        & -0.223        &  -0.168        & -0.087        & -0.309        & 0.198 \\
    \hline
    $[331]^{24}S$~ & $[331]^{22}D$~ & $[331]^{24}D$~ & $[322]^{22}P$~ & $[322]^{24}P$~ & $[322]^{26}P$~ & \\
    0.158         & 0.123         & 0.043         & -0.016        & 0.080          & 0.080 & \\
    \hline
    \hline
    \end{tabular}

    \vspace{1cm}
    
              $E^*=2.12 (1.9)$ MeV, $T=1/2, J=1/2$

    \begin{tabular}{llllllll}
    \hline
    \hline
    $[43]^{22}P$~ & $[421]^{22}P$~ & $[421]^{24}P$~ & $[421]^{24}D$~ & $[331]^{22}S$~ & $[331]^{24}D$~ & $[322]^{22}P$~ & $[322]^{24}P$ \\
    0.913        & 0.161         & -0.132        & -0.314        & 0.126          & -0.088        & 0.000         & 0.001        \\ 
    \hline
    \hline
    \end{tabular}

    \vspace{1cm}

             $E^*=5.02 (6.9)$ MeV, $T=1/2, J=3/2$

    \begin{tabular}{lllllll}
    \hline
    \hline   
    $[43]^{22}P$~ & $[43]^{22}D$~ & $[421]^{22}P$~ & $[421]^{24}P$~ & $[421]^{22}D$~ & $[421]^{24}D$~ & $[421]^{24}F$ \\
    -0.532       & 0.721        & -0.061        & 0.207         & 0.272         &  0.036         & 0.079 \\ 
    \hline
    $[331]^{24}S$~ & $[331]^{22}D$~ & $[331]^{24}D$~ & $[322]^{22}P$~ & $[322]^{24}P$~ & $[322]^{26}P$ & \\
    -0.166        & 0.021         & 0.155         & 0.048         & -0.039        & -0.111        & \\
    \hline
    \hline
    \end{tabular} 
    
  \end{center}
\end{table}

\section{The reaction model}
\label{model}

\begin{figure}
  \begin{center}
    \includegraphics[scale = 0.50]{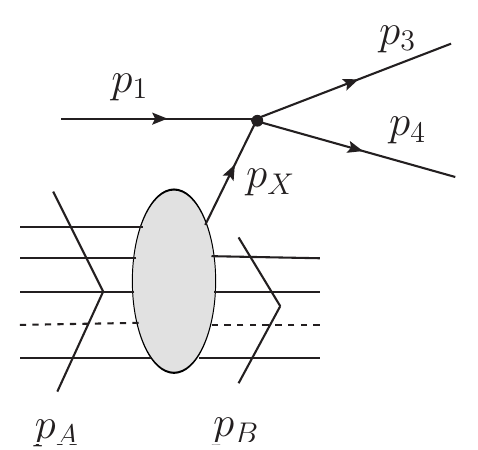}
  \end{center}
  \caption{\label{fig:diagr} The amplitude of the process $A(p,2p)B$. 
    The lines are marked with four-momenta of the particles:
    the initial ($p_A$) and final ($p_B$) nuclei, incident proton ($p_1$),
    struck proton ($p_X$), outgoing protons ($p_3$ and $p_4$).}
\end{figure}

In the Feynman diagram representation, the amplitude of the studied process is shown in Fig.~\ref{fig:diagr} which
gives the following invariant matrix element:
\begin{equation}
    M = M_{\rm el}(p_3,p_4,p_1) 
              \frac{i\Gamma_{A \to X B}(p_A,p_B)}{p_X^2-m^2+i\epsilon}~,  \label{M}
\end{equation}
where $M_{\rm el}(p_3,p_4,p_1)$ is the invariant matrix element of elastic $pp$ scattering amplitude, $\Gamma_{A \to X B}(p_A,p_B)$ is the nuclear virtual
decay vertex, and $m$ is the nucleon mass. The sum over all intermediate state quantum numbers is implicitly assumed in Eq.(\ref{M}).
For the residual nucleus $B$ on the mass shell, in the rest frame (r.f) of the initial nucleus $A$, the
decay vertex can be expressed as follows:
\begin{equation}
  \frac{i\Gamma_{A \to X B}(p_A,p_B)}{p_X^2-m^2+i\epsilon}
  = S_A^X \left(\frac{2 E_B m_A}{p_X^0}\right)^{1/2} (2\pi)^{3/2} \psi_{nl}^{m_l}(-\bvec{p}_X)~,  \label{Gamma_A}
\end{equation}
where $\psi_{nl}^{m_l}(-\bvec{p}_X)$ is the WF of the relative motion of the struck proton $X$ and the nucleus $B$
in momentum space
with $n$ being the HO main quantum number, $l$ -- the relative orbital momentum, and $m_l$ -- the magnetic
quantum number.
The WF is normalized as follows:
\begin{equation}
  \int d^3p_X |\psi_{nl}^{m_l}(-\bvec{p}_X)|^2 = 1~.   \label{normWF}
\end{equation}

The spectroscopic amplitude $S_A^X$ for the case when the ``cluster $X$'' is simply the nucleon  can be obtained from
a more general Eq.(\ref{S_A^X}) for $b=1, L_X=0, J_X=S_X=T_X=1/2, \Lambda=l, M_\Lambda=m_l, M_X=\sigma, M_{T_X}=\tau$ which gives:
\begin{eqnarray}
  \lefteqn{S_A^X([f] LS(J)T MM_T;[f_B] L_BS_B(J_B)T_B M_BM_{T_B};m_l,\sigma)} \nonumber \\
  &=& A^{1/2}  
   \sum_{J_0 M_0}
   \left\{\begin{array}{lll}
              L_B      & S_B & J_B \\
              l        & 1/2 & J_0 \\
              L        & S   & J
          \end{array}
   \right\} \sqrt{(2L+1)(2S+1)(2J_B+1)(2J_0+1)}  \nonumber \\
  && \times \langle AN[f] LST|(A-1)N_B[f_B] L_BS_BT_B; nl \rangle \nonumber \\
  && \times  (J_B M_B J_0 M_0 | J M)\, (l m_l \frac{1}{2} \sigma | J_0 M_0)\,
            (T_B M_{T_B} \frac{1}{2} \tau | T M_T)~,      \label{S_A}
\end{eqnarray}
where the state of the struck nucleon $X$ is determined by the spin, $\sigma$, and isospin,  $\tau$,  projections.
\footnote{For clarity, in Eq.(\ref{S_A}) and below in this section we explicitly include the arguments of the spectroscopic factor
and FPC. The Elliott symbol is redundant for transitions involving only $p$-shell nucleons since it is related to the Young scheme
$[f_1f_2f_3]$ as $(\lambda\mu)=(f_1-f_2,f_2-f_3)$.}
The numbers of oscillator quanta satisfy the sum rule $N=N_B+n$.

The relation between the one-particle FPC of the TISM (the term in the angular brackets in Eq.(\ref{S_A}))
and the one-particle FPC of the conventional shell model also follows from a more general Eq.(\ref{FPC_TISM_vs_usual}).
For the WF of relative $X-B$ motion with quantum numbers $n=l=1$ the cluster coefficient is equal to unity and,
thus, the following simple formula holds (see also Ref.~\cite{Kurdyumov:1970qkz}):
\begin{eqnarray}
  && \langle A N[f] LST | (A-1) N_B[f_B] L_BS_BT_B; 1 1 \rangle \nonumber \\
  && = - \left(\frac{A-4}{A-1}\right)^{1/2}
         \langle p^{A-4}[f] LST | p^{A-5}[f_B] L_BS_BT_B \rangle~. \label{FPCusual}
\end{eqnarray}
The one-particle FPC of the conventional shell model can be calculated in a standard way from the tables of Ref.~\cite{Jahn51}
taking into account the correction of phases of some orbital WFs as mentioned in the footnote of Ref.~\cite{Elliott53}.

Equation (\ref{Gamma_A}) assumes transition from a TISM state of the nucleus $A$ to a TISM state of the nucleus $B$.
Eigenstates of a realistic nuclear Hamiltonian should be the superposition of TISM states. This means that the actual spectroscopic amplitude
of the transition between the physical states of the nuclei $A$ and $B$ is obtained by a weighted sum,
\begin{eqnarray}
 && \sum_{i,j} \alpha_i^{A,JT} \alpha_j^{A-1,J_BT_B}  S_A^X([f_i] L_iS_i(J)T MM_T;[f_j] L_jS_j(J_B)T_B M_BM_{T_B};m_l,\sigma) \nonumber \\
 && \equiv S_A^X(JTMM_T;J_BT_BM_BM_{T_B};m_l,\sigma)~, \label{sum_spec}
\end{eqnarray}
where the amplitudes $\alpha$ enter the decomposition of the physical states of the nuclei $A$ and $B$, see Eq.(\ref{GTISM}).

So far we discussed the case of IA neglecting ISI/FSI.
We will now include the ISI/FSI in the eikonal approximation, similar to Ref.~\cite{Larionov:2023cpe}.
This is reached by replacing  in Eq.~(\ref{Gamma_A})
\begin{equation}
    (2\pi)^{3/2} \psi_{nl}^{m_l}(-\bvec{p}_{X})
    \to  \int d^3r \mbox{e}^{-i\bvec{p}_{X}\bvec{r}} \psi_{nl}^{m_l}(-\bvec{r})
    F_1(\bvec{r}) F_3(\bvec{r})  F_4(\bvec{r})
    \equiv (2\pi)^{3/2} \tilde{\psi}_{nl}^{m_l}(-\bvec{p}_{X})~,         \label{FSI_corr}
\end{equation}
where $\bvec{r}=\bvec{R}_X-\bvec{R}_B$ is the relative position vector of the struck proton and c.m. of the residual nucleus $B$.
Thus, Eq.(\ref{FSI_corr}) takes into account nuclear absorption introduced via factors 
\begin{equation}
  F_j(\bvec{r}) = \exp\left(-\frac{i}{v_j} \int\limits_{-\infty}^0 d\eta\, U_j(\bvec{r} \pm \hat{\bvec{p}}_j \eta) \right)~, \label{F_j}
\end{equation}
where $v_j=p_j/E_j$ is the $j$-th particle velocity in the rest frame (r.f.) of the nucleus $B$,
$\hat{\bvec{p}_j} \equiv \bvec{p}_j/p_j$, and $U_j(\bvec{r})$ is the optical potential.
In Eq.(\ref{F_j}), the integral is taken along the trajectory of the $j$-th particle that corresponds to the ``+'' sign for $j=1$
(incoming proton) and ``-'' sign for $j=3,4$ (outgoing protons).
At relativistic energies, in good approximation, the latter can be expressed as follows: 
\begin{equation}
   U_j(\bvec{r}) = -\frac{i}{2} v_j \sigma_{NN}(p_j) \rho(\bvec{r})~, \label{U_j}
\end{equation}  
where $\sigma_{NN}(p_j)$ is the total $NN$ cross section, and $\rho(\bvec{r})$ is the nucleon number density of the nucleus $B$ in the point $\bvec{r}$.
The resulting absorption factors are then essentially similar to those in the Glauber approximation \cite{VanOvermeire:2006dr}.

We will consider the case of unpolarized particles and, thus, the matrix element modulus squared should be averaged over spin magnetic
quantum numbers of the initial particles and summed over those of final particles:
\begin{eqnarray}
  \overline{|M|^2} &\equiv&  \frac{1}{2(2J+1)} \sum_{\sigma_1,\sigma_3,\sigma_4,M,M_B}  |M|^2 \nonumber \\
  &=& \frac{1}{2(2J+1)} \sum_{\sigma_1,\sigma_3,\sigma_4,M,M_B} \sum_{\sigma,\sigma^\prime} \sum_{m_l,m_l^\prime} M_{\rm el}(p_3,p_4,p_1;\sigma_3,\sigma_4,\sigma_1,\sigma)
  M_{\rm el}^*(p_3,p_4,p_1;\sigma_3,\sigma_4,\sigma_1,\sigma^\prime) \nonumber \\
  && \times \frac{2 E_B m_A}{p_X^0}  (2\pi)^3 \tilde{\psi}_{11}^{m_l}(-\bvec{p}_X) \tilde{\psi}_{11}^{m_l^\prime *}(-\bvec{p}_X) \nonumber \\
  && \times S_A^X(JTMM_T;J_BT_BM_BM_{T_B};m_l,\sigma)  S_A^{X *}(JTMM_T;J_BT_BM_BM_{T_B};m_l^\prime,\sigma^\prime)~,   \label{M^2}
\end{eqnarray}
where we explicitly included summations over intermediate state quantum numbers $\sigma,\sigma^\prime,m_l,m_l^\prime$. To simplify Eq.(\ref{M^2}), we, first,
neglect the interference terms with $\sigma^\prime \neq \sigma$ and replace
\begin{eqnarray}
  && \frac{1}{2} \sum_{\sigma_1,\sigma_3,\sigma_4}  |M_{\rm el}(p_3,p_4,p_1;\sigma_3,\sigma_4,\sigma_1,\sigma)|^2 \nonumber \\
  &&  \to \overline{|M_{\rm el}(p_3,p_4,p_1)|^2} \equiv \frac{1}{4} \sum_{\sigma_1,\sigma_3,\sigma_4,\sigma} |M_{\rm el}(p_3,p_4,p_1;\sigma_3,\sigma_4,\sigma_1,\sigma)|^2
  \label{M_el^2}~.
\end{eqnarray}
Then, after somewhat lengthy but straightforward derivation we come to the following expression:
\begin{eqnarray}
  \overline{|M|^2} &=& \overline{|M_{\rm el}(p_3,p_4,p_1)|^2}\,\frac{2 E_B m_A}{p_X^0} (2\pi)^3\, \overline{|\tilde{\psi}_{11}(-\bvec{p}_X)|^2}\,
  S~,    \label{M^2_fin}
\end{eqnarray}
where
\begin{equation}
  \overline{|\tilde{\psi}_{11}(-\bvec{p}_X)|^2} \equiv  \frac{1}{3} \sum_{m_l=-1}^1 |\tilde{\psi}_{11}^{m_l}(-\bvec{p}_X)|^2~.    \label{psi_11^2}
\end{equation}
The spectroscopic factor in Eq.(\ref{M^2_fin}) is expressed as follows:
\begin{eqnarray}
  S &=& A\, (T_B M_{T_B} \frac{1}{2} \tau | T M_T)^2\, (2J_B+1)\, \sum_{J_0} (2J_0+1)\, |\sum_{i,j} \alpha_i^{A,JT} \alpha_j^{A-1,J_BT_B} 
    \left\{\begin{array}{lll}
              L_j      & S_j  & J_B \\
              1        & 1/2  & J_0 \\
              L_i      & S_i  & J
          \end{array}
   \right\}    \nonumber \\
   && \times \sqrt{(2L_i+1)(2S_i+1)}\,  \langle AN[f_i] L_iS_iT|(A-1)N_B[f_j] L_jS_jT_B; 11 \rangle |^2~.   \label{S}
\end{eqnarray}

Equation (\ref{psi_11^2}) for the modulus squared of the ISI/FSI-corrected WF  is quite involved
but can be simplified.
Substituting Eq.(\ref{FSI_corr}) in Eq.(\ref{psi_11^2}) we have:
\begin{eqnarray}
  \overline{|\tilde{\psi}_{11}(-\bvec{p}_X)|^2} &=& \frac{1}{3(2\pi)^3}  \sum_{m_l} \int d^3r \int d^3r^\prime
  \mbox{e}^{-i\bvec{p}_{X}(\bvec{r}-\bvec{r}^\prime)}
           \psi_{11}^{m_l}(-\bvec{r}) \psi_{11}^{m_l *}(-\bvec{r}^\prime)
           F_{\rm abs}(\bvec{r})  F_{\rm abs}(\bvec{r}^\prime)  \nonumber \\
 &\simeq&  \frac{1}{(2\pi)^3} \int d^3R\, f_{11}(-\bvec{R},-\bvec{p}_{X})  F_{\rm abs}^2(\bvec{R})~,    \label{psi_11^2_sim}
\end{eqnarray}
where $F_{\rm abs}(\bvec{r}) \equiv F_1(\bvec{r}) F_3(\bvec{r})  F_4(\bvec{r})$ is the absorption factor, and
\begin{equation}
  f_{11}(-\bvec{R},-\bvec{p}_{X}) \equiv  \frac{1}{3} \sum_{m_l} \int d^3\xi \mbox{e}^{-i\bvec{p}_{X}\bvec{\xi}}\, \psi_{11}^{m_l}(-\bvec{R}-\bvec{\xi}/2)
  \psi_{11}^{m_l\, *}(-\bvec{R}+\bvec{\xi}/2)    \label{WignerFunction}
\end{equation}
is a Wigner function. 
In Eq.(\ref{psi_11^2_sim}), in the last step, we introduced variables $\bvec{R} \equiv (\bvec{r}+\bvec{r}^\prime)/2$,
$\bvec{\xi} \equiv \bvec{r}-\bvec{r}^\prime$ and approximately set $\bvec{\xi}=0$ in the product of absorption factors
$F_{\rm abs}(\bvec{R}+\bvec{\xi}/2) F_{\rm abs}(\bvec{R}-\bvec{\xi}/2)$ (see Ref.~\cite{Larionov:2023cpe} for discussion
of validity of this approximation).

The TISM WF of the relative $X-B$ motion
is
\begin{equation}
  \psi_{11}^{m_l}(-\bvec{R}) = \left(\frac{8}{3\pi^{1/2}R_0^5}\right)^{1/2} R\, \mbox{e}^{-R^2/2R_0^2}\, Y_{1m_l}(-\hat{\bvec{R}})~, \label{psi_11}
\end{equation}
where $R_0 = r_0 [A/(A-1)]^{1/2}$.
The Wigner function (\ref{WignerFunction}) can be then easily calculated:
\begin{equation}
  f_{11}(-\bvec{R},-\bvec{p}_{X}) = 8\, \mbox{e}^{-(R^2+p_X^2 R_0^4)/R_0^2} \left[ \frac{2}{3R_0^2}(R^2+p_X^2 R_0^4) -1 \right]~.   \label{f_11}
\end{equation}
This expression can be used in Eq.(\ref{psi_11^2_sim}) for numerical calculations of $\overline{|\tilde{\psi}_{11}(-\bvec{p}_X)|^2}$
taking into account ISI/FSI. In the case of IA, one recovers an analytical formula:
\begin{equation}
  \overline{|\psi_{11}(-\bvec{p}_X)|^2} = \frac{2 R_0^5}{3\pi^{3/2}}\, p_X^2\, \mbox{e}^{-p_X^2 R_0^2}~.   \label{psi_11^2_woAbs}
\end{equation}

The invariant matrix element of elastic $pp$ scattering is related to the differential cross section by a standard formula:
\begin{equation}
  \frac{d\sigma_{\rm el}}{dt} = \frac{\overline{|M_{\rm el}(t,s)|^2}}{64 \pi I_{pp}^2}~,   \label{dsig_el/dt}
\end{equation}
where $I_{pp}=\sqrt{(s/4-m^2)s}$ is the flux factor, $s=(p_3+p_4)^2$, $t=\max\{(p_1-p_3)^2,(p_1-p_4)^2\}$.
By using the high-energy parameterization $d\sigma_{\rm el}/dt \propto \mbox{e}^{bt}$
one obtains the following relation:
\begin{equation}
   \overline{|M_{\rm el}(t,s)|^2} = 64 \pi I_{pp}^2\, \frac{b\sigma_{\rm el}}{1-\mbox{e}^{bt_0}}\, \mbox{e}^{bt}~,   \label{M_el^2_par}
\end{equation}
where $t_0=-2(s/4-m^2)$. The experimental integrated elastic $pp$ cross section, $\sigma_{\rm el}$, and the slope parameter, $b$, are conveniently
parameterized in Ref.~\cite{Cugnon:1996kh} as functions of the beam momentum, $p_{\rm lab}=I_{pp}/m$, for $p_{\rm lab} \ltsim 5-6$ GeV/c.

For the calculation of the optical potential, Eq.(\ref{U_j}), one has to specify the total $pN$ cross section and the nucleon density distribution.
We apply the proton/neutron-number-weighted formula
\begin{equation}
  \sigma_{pN} = [\sigma_{pp} Z_B + \sigma_{pn} (A_B-Z_B)]/A_B~,  \label{sigma_pN}
\end{equation}
where $\sigma_{pp}$ and $\sigma_{pn}$ are, respectively, the total $pp$ and $pn$ cross sections in the parameterizations of Ref.~\cite{Cugnon:1996kh}
that provide good fits of available experimental data at $p_{\rm lab} \ltsim 3-5$ GeV/c. 
$A_B=A-1$ and $Z_B=Z-1$ are, respectively, the mass and charge numbers of the residual nucleus $B$.
The nucleon density distribution in the nucleus with $A_B$ nucleons in the $s^4p^{A_B-4}$ configuration
is described by the conventional HO shell model formula
\begin{equation}
   \rho(r) = \frac{4}{r_0^3 \pi^{3/2}} \left[1 + \frac{A_B-4}{6} \left(\frac{r}{r_0}\right)^2\right] \mbox{e}^{-r^2/r_0^2}~.    \label{rho}
\end{equation}

The fully differential cross section of the process $A(p,2p)B$ (see Fig.~\ref{fig:diagr} for notation) is expressed as follows:
\begin{equation}
  d\sigma_{1A \to 34B} = \frac{(2\pi)^4 \overline{|M|^2}}{4I_{pA}} \delta^{(4)}(p_1+p_A-p_3-p_4-p_B) \frac{d^3p_3}{(2\pi)^32E_3} \frac{d^3p_4}{(2\pi)^32E_4}
  \frac{d^3p_B}{(2\pi)^32E_B}~,     \label{dsigma_1A34B}
\end{equation}
where $I_{pA}=p_{\rm beam} m$ is the flux factor, $p_{\rm beam}$ is the momentum of the nucleus $A$ in the r.f. of the proton 1 which will be called ``laboratory frame'' below.. 
The experiment of Ref.~\cite{Panin:2016div} has been performed at $p_{\rm beam}/A=0.951$ GeV/c where the $pp$ elastic cross section is almost isotropic
in the effective region of integration over transferred rescattering  momentum in the c.m. frame as follows from the used here parameterization \cite{Cugnon:1996kh}.

Thus, it is convenient to perform the integration over invariants $d^3p_3/E_3$ and $d^3p_4/E_4$ in the c.m. frame of the protons 3 and 4. On the other hand, the
matrix element is proportional to the WF of the relative $X-B$ motion which suppresses large absolute values of the momentum $\bvec{p}_B$ of the residual nucleus in the r.f. of the
nucleus $A$. Thus, the integration over invariant $d^3p_B/E_B$ is reasonable to perform in the  r.f. of the nucleus $A$. As a result, we come to the following expression for the
integrated cross section:
\begin{equation}
  \sigma_{1A \to 34B} = \frac{1}{32(2\pi)^5p_{\rm beam}m} \int \frac{d^3p_B}{\sqrt{p_B^2+m_B^2}} \int \frac{p_3 d\Omega_3}{2\sqrt{p_3^2+m^2}} \overline{|M|^2}~, \label{sigma_1A34B}
\end{equation}
where the momentum $\bvec{p}_B$ is defined in the r.f. of the nucleus $A$ while the momentum $\bvec{p}_3$ and the corresponding solid angle element $d\Omega_3$ -- in the c.m.
frame of the protons 3 and 4. Due to identity of the differential cross section with respect to the interchange of momenta of the 3-d and 4-th protons, the integration
over $d\Omega_3$ should be performed over the solid angle hemisphere of $2\pi$ (the orientation of the hemisphere does not play a role). Due to rotational symmetry about
the beam axis, it is also possible to reduce the integration order by writing in the spherical coordinates with $z$-axis along the 1-st proton momentum in the r.f. of $A$
$d^3p_B= p_B^2 dp_B 2\pi d\Theta_B$ and perform all integrations in Eq.(\ref{sigma_1A34B}) with arbitrarily fixed value of the azimuthal angle $\phi_B$. 
Finally, the differential cross sections, $d\sigma_{1A \to 34B}/dx$, where $x$ is any kinematic
variable determined  by the  momenta ${\bf p}_3$ and  ${\bf p}_B$,
are evaluated by multiplying the integrand of Eq.(\ref{sigma_1A34B})
by the factor $\delta[x-x(\bvec{p}_B,\bvec{p}_3)]$.

\section{Results}
\label{results}

The integrated cross sections are listed in Table~\ref{tab:sigma}. One can see from this Table that absorption
reduces all partial cross section by a factor of 5.1 but does not change the ratios between the partial cross sections
for different states of $^{11}$B. The experimental total cross section is reproduced by full calculation surprisingly well, with accuracy of about $3\%$.
The strong dominance of $^{11}$B production
in the ground state is also correctly reproduced. Discrepancies for excited states are quite large.
However, this is still satisfactory given the fact that we did not introduce any additional model parameters (like phenomenological spectroscopic factors, see discussion section)
to tune our calculations.
\begin{table}[htb]
  \caption{\label{tab:sigma} Integrated cross sections of the process $^{12}\mbox{C}(p,2p)^{11}\mbox{B}$ with 400 MeV/nucleon $^{12}$C beam
    for the ground state and two excited states of the residual nucleus $^{11}\mbox{B}$. Listed are the results of full calculations
    (including absorption), calculations in the IA, and the spectroscopic factors calculated using Eq.(\ref{S}). 
    Experimental data are from Ref.~\cite{Panin:2016div}. Total errors are given in parentheses.}
    \begin{center}
    \begin{tabular}{llllll}
    \hline
    \hline
    $E^*$ (MeV)~~~ & $J^\pi$~~~ &   $\sigma_{\rm exp}$ (mb)~~~        &  $\sigma_{\rm full}$ (mb)~~~ & $\sigma_{\rm IA}$ (mb)~~~ &  $S$ \\
    \hline
    0.0 (G.S.)     & $3/2^-$   &   $15.8(18)$                      &   12.3                     &  62.6                   & 2.82  \\
    2.12           & $1/2^-$   &   $1.9(2)$                        &   2.9                      &  14.9                   & 0.67  \\
    5.02           & $3/2^-$   &   $1.5(2)$                        &   3.4                      &  17.5                   & 0.79  \\
    \hline
    Total:         &           &   $19.2(3)$                       &   18.6                     &  95.0                   & 4.28  \\
    \hline
    \hline
    \end{tabular}
  \end{center} 
\end{table}

\begin{figure}
  \begin{center}
    \includegraphics[scale = 0.60]{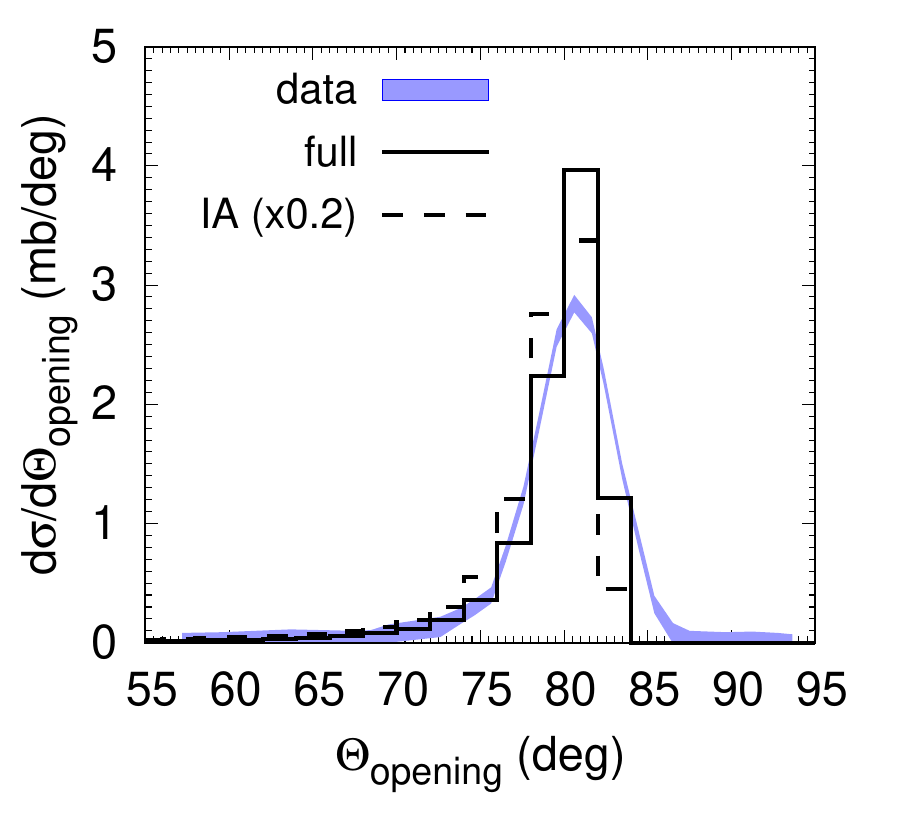}
  \end{center}
  \caption{\label{fig:Theta} The distribution of opening angle
    between outgoing protons in the laboratory frame for the process $^{12}\mbox{C}(p,2p)^{11}\mbox{B}$ at 400 MeV/nucleon.
    Solid and dashed histograms show, respectively, the full calculation and the calculation without absorption scaled by a factor of 0.2.
    The band represents experimental data from Ref.~\cite{Panin:2016div}.}
\end{figure}

Fig.~\ref{fig:Theta} shows the distribution of opening angle,
$\Theta_{\rm opening}= \arccos(\bvec{p}_3\bvec{p}_4/p_3p_4)$,
between outgoing protons in the laboratory frame.
The full calculation correctly describes the peak position at $80\degree$ and the distribution at smaller angles,
although gives a sharper peak and steeper fall-off at larger angles. The full calculation is slightly shifted
to larger opening angles as compared to the calculation in the IA. 

\begin{figure}
  \begin{center}
    \includegraphics[scale = 0.60]{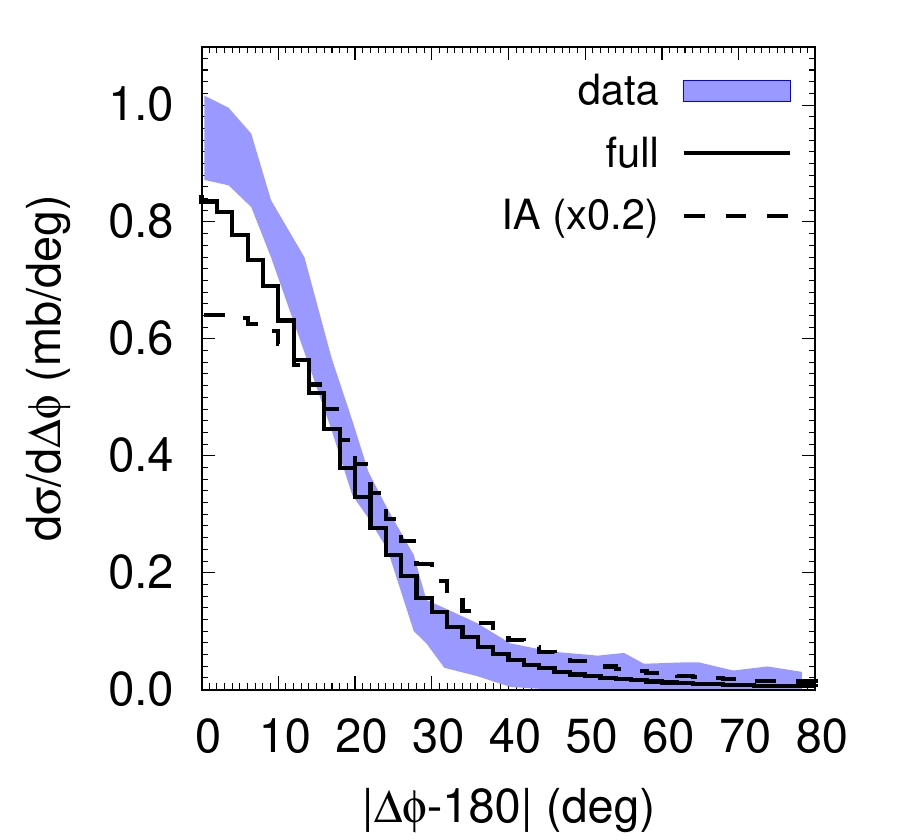}
  \end{center}
  \caption{\label{fig:phi} The distribution of the relative azimuthal angle between transverse momenta of
    outgoing protons for $^{12}\mbox{C}(p,2p)^{11}\mbox{B}$ at 400 MeV/nucleon. Line notations are the same as in Fig.~\ref{fig:Theta}.
    The band represents experimental data from Ref.~\cite{Panin:2016div}.}
\end{figure}

Fig.~\ref{fig:phi} shows  the distribution of relative azimuthal angle, $\Delta\phi= \arccos(\bvec{p}_{3t}\bvec{p}_{4t}/p_{3t}p_{4t})$,
between the transverse momenta $\bvec{p}_{3t}$ and $\bvec{p}_{4t}$ of outgoing protons.
The absorptive ISI/FSI suppress the yield at large deviations of $\Delta\phi$ from $180\degree$
leading to a sharper peak at $\Delta\phi=180\degree$ and better agreement with experiment. 

\begin{figure}
  \begin{center}
    \includegraphics[scale = 0.60]{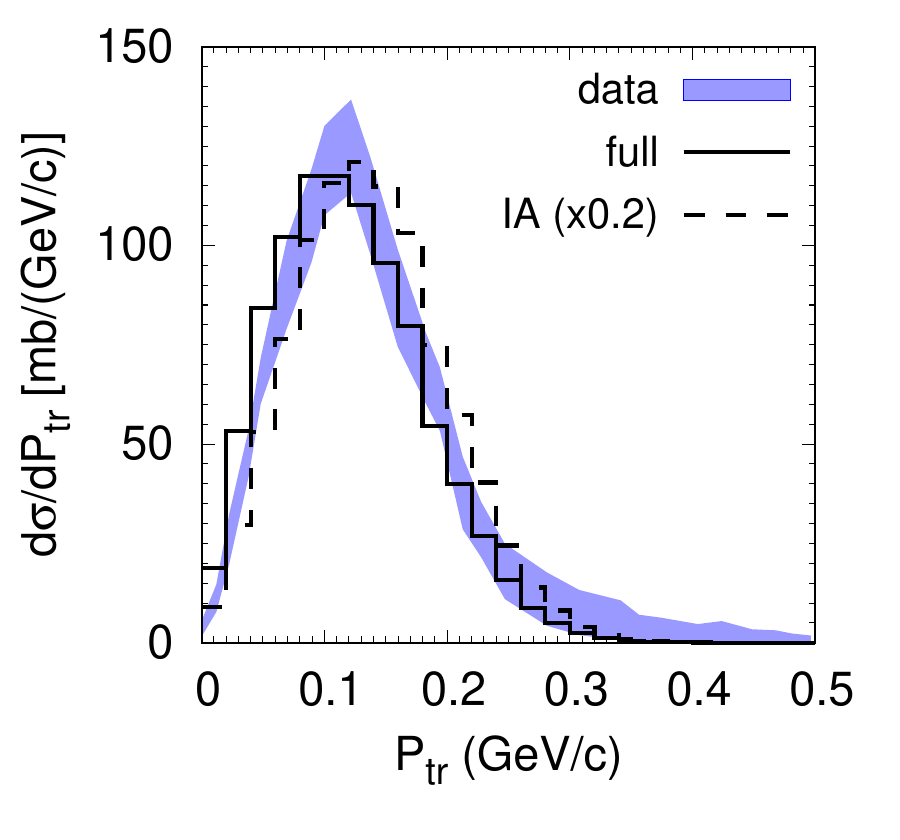}
  \end{center}
  \caption{\label{fig:ptr} The distribution of the transverse momentum of the residual nucleus for $^{12}\mbox{C}(p,2p)^{11}\mbox{B}$ at 400 MeV/nucleon.
   Line notations are the same as in Fig.~\ref{fig:Theta}. The band represents experimental data from Ref.~\cite{Panin:2016div}.}
\end{figure}

\begin{figure}
  \begin{center}
    \includegraphics[scale = 0.60]{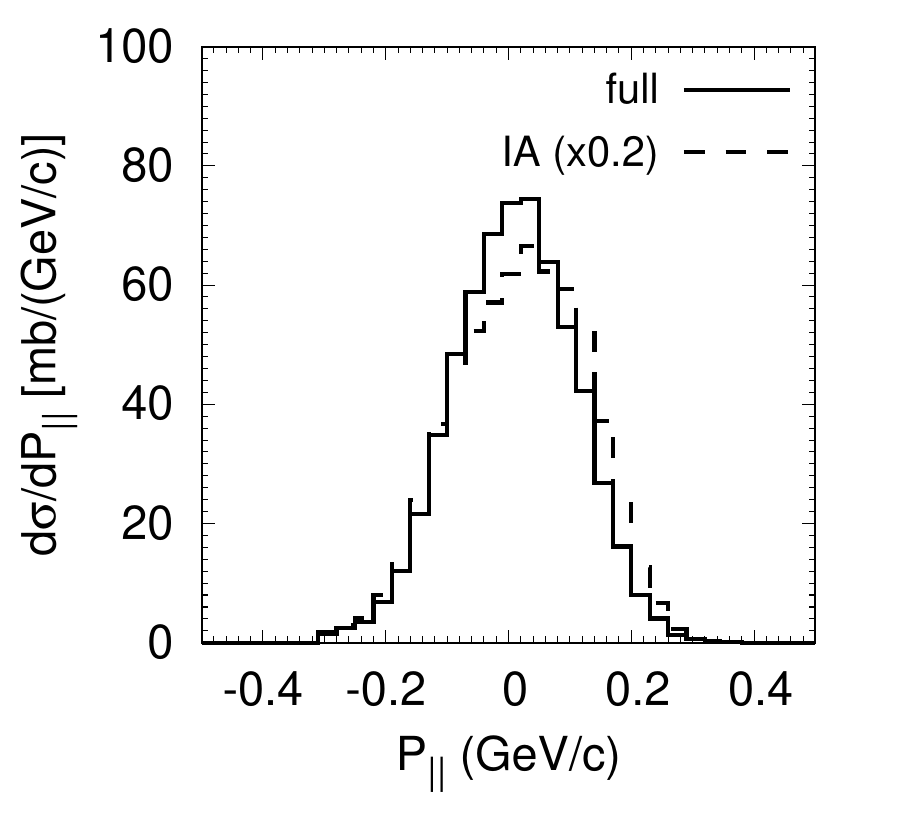}
  \end{center}
  \caption{\label{fig:plong} The distribution of the longitudinal momentum of the residual nucleus
    in the r.f. of $^{12}$C for $^{12}\mbox{C}(p,2p)^{11}\mbox{B}$ at 400 MeV/nucleon.
    Positive values of $P_\parallel$ correspond to the direction of the momentum of the incoming proton. 
    Line notations are the same as in Fig.~\ref{fig:Theta}.}
\end{figure}

\begin{figure}
  \begin{center}
    \includegraphics[scale = 0.60]{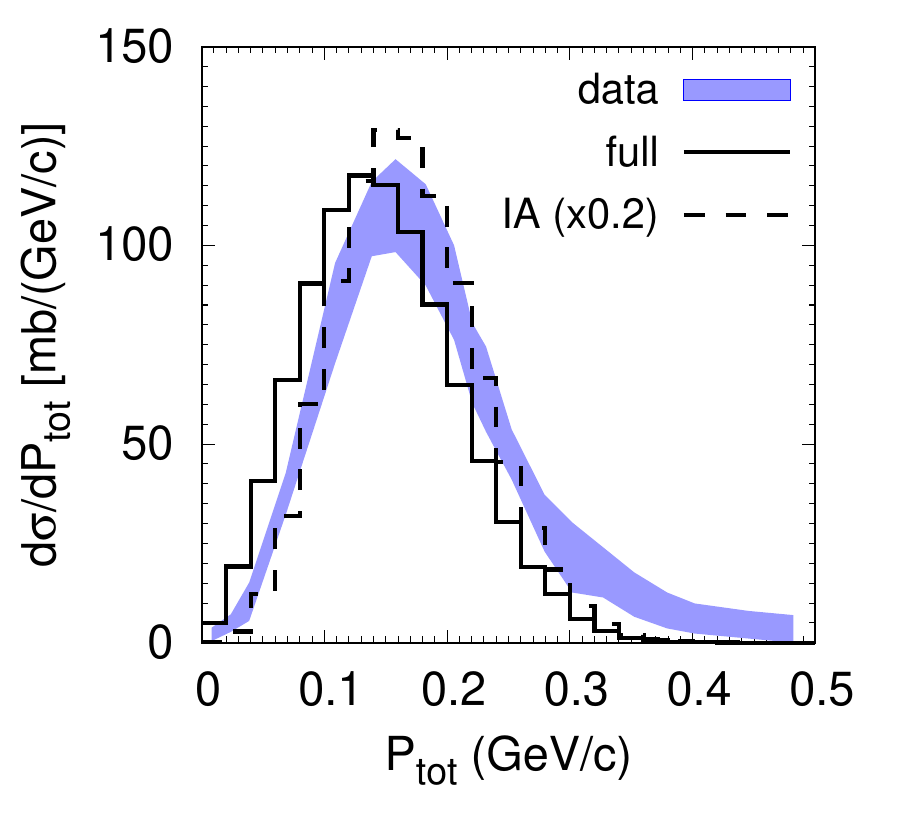}
  \end{center}
  \caption{\label{fig:ptot} The distribution of the total momentum of the residual nucleus in the r.f. of $^{12}$C for $^{12}\mbox{C}(p,2p)^{11}\mbox{B}$ at 400 MeV/nucleon.
   Line notations are the same as in Fig.~\ref{fig:Theta}. The band represents experimental data from Ref.~\cite{Panin:2016div}.}
\end{figure}

Figs.~\ref{fig:ptr},\ref{fig:plong} and \ref{fig:ptot} show, respectively, the transverse, longitudinal, and total momentum distributions
of the residual nucleus. Absorption leads to the depletion of the yield at large transverse and total momenta shifting the maxima of the
$P_{\rm tr}$- and $P_{\rm tot}$ distributions to smaller momenta. This can be understood as follows. In the presence of absorption, the main
contribution to the integral in Eq.(\ref{psi_11^2_sim}) comes from nuclear periphery (surface ring), since the absorption factor
suppresses the integrand deeply inside the nucleus. If the transverse momentum of the residual nucleus is small, then the absorption is in average
smaller because both outgoing protons may have small transverse momenta balancing each other and their trajectories avoid the bulk of the nuclear medium.
If the transverse momentum of the residual nucleus is large, then at least one of the outgoing protons will have large transverse momentum and, thus,
its trajectory will pass through the bulk of the nuclear medium with a larger probability which makes absorption stronger.
This is also in-line with stronger absorption at larger deviations of $\Delta\phi$ from $180\degree$ (Fig.~\ref{fig:phi}). 

A particular form of the Wigner density for the $n=l=1$ valence nucleon, Eq.(\ref{f_11}), acts in the same direction.
At small values of $R$ and $p_X (=P_{\rm tot})$, the Wigner density becomes negative. It means that absorption should then lead to the enhancement of production
which is visible at small values of $P_{\rm tot}$ in Fig.~\ref{fig:ptot}.

In the IA calculation, the longitudinal momentum distribution (Fig.~\ref{fig:plong}) is shifted to positive $P_{\rm ||}$. This corresponds to the struck proton $X$
moving opposite to the incoming proton 1 in the r.f. of $^{12}$C giving a larger two-body phase space volume of the protons 3 and 4 (the term $\propto p_3/\sqrt{p_3^2+m^2}$
in Eq.(\ref{sigma_1A34B})). However, with absorption, the $P_{\rm ||}$ distribution becomes almost symmetric with respect to the change $P_{\rm ||} \to -P_{\rm ||}$.
This is a consequence of larger average transverse momenta of the 3-d and 4-th protons at $P_{\rm ||} > 0$ leading to their stronger absorption.
This observation also explains the stronger absorption at smaller opening angles (Fig.~\ref{fig:Theta}). 

\section{Discussion}
\label{discuss}

Table~\ref{tab:S} contains results of several other calculations of the spectroscopic factors for the separation of a nucleon from $^{12}$C
in comparison with our results.
\begin{table}[htb]
  \caption{\label{tab:S} Spectroscopic factors for the $^{12}\mbox{C} \to p ^{11}\mbox{B} (n ^{11}\mbox{C})$ process calculated in different theoretical models.} 
    \begin{center}
    \begin{tabular}{llllll}
    \hline
    \hline
    $E^*$ (MeV)~~~ & $J^\pi$~~~~~  &   \cite{Cohen:1967zzb}~~~&  \cite{Singh:1973fqv}~~ & \cite{Li:2022zao}~~ &  this work  \\
    \hline
    0.0 (G.S.)     & $3/2^-$          &   2.85                   &  3.27                 &  2.50   & 2.82  \\
    2.12           & $1/2^-$          &   0.75                   &  0.60                 &  0.48  & 0.67  \\
    5.02           & $3/2^-$          &   0.38                   &  0.12                 &      & 0.79  \\
    \hline
    Total:         &                   &   3.98                  &  3.99                 &  2.98 & 4.28  \\
    \hline
    \hline
    \end{tabular}
  \end{center} 
\end{table}
The approach of Ref.~\cite{Cohen:1967zzb} is based on the FPCs of the conventional shell model but in other aspects
is quite close to the intermediate coupling model of Refs.~\cite{Balashov:1964,Boyarkina73} applied in our work.
In Ref.~\cite{Singh:1973fqv}, the WFs of deformed HO shell model were used without residual interaction
taking $k=0$ for $^{12}$C and $k=1/2,3/2$ for $^{11}$B. 
(In the spherical HO model this corresponds to considering only the Young scheme $[44]$ for $^{12}$C and $[43]$ for $^{11}$B.)
In Ref.~\cite{Li:2022zao}, the  no-core shell model was employed in the HO basis with c.m. correction, although
the authors state that the dependence of their results on the chosen basis is small. According to Ref.~\cite{Rodkin:2021chh},
however, the ``old'' definition of the spectroscopic amplitude (c.f. our Eq.(\ref{S_A})) relies on
the non-normalized WF of the final state and, thus, should be corrected.
\footnote{It was pointed out in Ref.~\cite{Rodkin:2021chh} that ``the numerical differences in the
results of calculations employing the “old” and “new” definitions are usually not large for single-nucleon channels, in
contrast to cluster channels''. Thus, we do not expect much influence of this correction on our results.}

Comparison with experimental data for the QE $(p,pN)$ processes depends not only on the spectroscopic factors
but also on the ISI/FSI used. In phenomenological DWIA approaches \cite{Devins:1979,Aumann:2013hga}, the spectroscopic factor is used as a free parameter
to fit experimental cross sections for some fixed ISI/FSI.  The latter  includes in-medium effects due to Pauli blocking
of $NN$ scattering \cite{Bertulani:2010kk} (i.e. the antisymmetrization of the full WF of the scattered nucleon and residual nucleus)
which can be effectively described by in-medium reduced $NN$ cross sections. It was shown \cite{Bertulani:2010kk}, that in heavy-ion induced
stripping reactions on nuclear targets at $E_{\rm lab}=5-300$ MeV/nucleon, the in-medium effects lead to about 10\% change in the nucleon knockout cross sections
and momentum distributions. In $(p,pN)$ processes, the in-medium effects are expected to be of the same order or smaller.
The in-medium effects should decrease with increasing beam energy rendering Glauber model more natural at $E_{\rm lab} > 1$ GeV/nucleon
\cite{VanOvermeire:2006dr}. Thus, our description of ISI/FSI within Glauber model should be taken with some reservations
for possible in-medium corrections.

\section{Summary}
\label{summary}

Based on the TISM, we developed the model for description of fully exclusive $A(p,pp)(A-1)$ reactions at intermediate relativistic energies.
The model allows to calculate spectroscopic factors directly from the overlap integral of the WFs.
Having in mind future model applications at NICA and FAIR energies, we restricted ourselves to the Glauber model description of the ISI/FSI.
As a test case, the model was applied to the reaction $^{12}\mbox{C}(p,2p)^{11}\mbox{B}$ at 400 MeV/nucleon 
measured at GSI \cite{Panin:2016div} in the inverse kinematics.

The model slightly underestimates the measured integrated cross section for $^{11}\mbox{B}$ $3/2^-$ ground state,
but overpredicts the integrated cross sections for the two excited $1/2^-$ and $3/2^-$ states.
The total integrated cross section for the production of all three $^{11}\mbox{B}$ states is well reproduced, however.

The distributions of the outgoing proton pair in the opening angle and relative azimuthal angle,
as well as the momentum distributions of the residual nucleus, are reproduced reasonably well.
Some deficiency in the production of high-momentum residual nuclei can be attributed to the longitudinal momenta mostly
and is probably due to the limited HO WF basis.

Last but not least, the present calculation also puts on a firm ground our previous study of the  $^{12}\mbox{C}(p,2pN_s)^{10}\mbox{A}$ exclusive reactions
at 48 GeV/c \cite{Larionov:2023cpe}, where a similar approach has been used.

\begin{acknowledgments}
The authors are grateful to Prof. Eliezer Piasetzky who proposed to apply  for  the $^{12}\mbox{C}(p,2p)^{11}\mbox{B}$ reaction the same TISM 
setup as for the $^{12}\mbox{C}(p,2pn_s)^{10}\mbox{B}$ reaction studied in Ref.~\cite{Larionov:2023cpe}.
\end{acknowledgments}

\bibliography{qe}

\newpage

\appendix

\section{Relation of the TISM FPCs and usual FPCs}
\label{FPC}

In this Appendix, we denote $\psi_{NL}(\bvec{R}_i)$ the HO WF of the c.m. motion of the nucleus $i=A,B,X$ with the
oscillator quantum number $N$ and orbital angular momentum $L$. For brevity, the magnetic quantum number is suppressed.  

The WF of the nucleus $A$ in the usual HO shell model can be represented as the antisymmetrized product
of the WFs of $s$- and $p$-shell nucleons:
\begin{eqnarray}
  && |s^4 p^{A-4} \alpha LST (x_1,x_2,\ldots,x_{A}) \rangle =
  \left(\begin{array}{c}
    A \\
    4
  \end{array}\right)^{-1/2} \sum (-1)^{\eta_{i_1,\ldots,i_A}} \psi_s(x_{i_1},x_{i_2},x_{i_3},x_{i_4}) \nonumber \\
  && \times  |p^{A-4} \alpha LST (x_{i_5},\ldots,x_{i_{A}})\rangle~,  \label{WF_A}   
\end{eqnarray}
where $x_i \equiv (\bvec{r}_i,\sigma_i,\tau_i)$ denote the position, $\bvec{r}_i$, spin, $\sigma_i$, and isospin, $\tau_i$,
variable of the $i$-th nucleon. The sum is taken over all possible sets of integer numbers $\{i_1,i_2,i_3,i_4\}$ and $\{i_5,\ldots,i_A\}$
selected from the set $\{1,2,\ldots,A\}$. The increasing order of the numbers is assumed in both sets.
$\eta_{i_1,\ldots,i_A}$ is the parity of the corresponding permutation.
The integrals 
$\langle s^4 | s^4 \rangle$ and $\langle p^{A-4} | p^{A-4} \rangle$ are assumed to be equal to one.
Due to the orthogonality of the single-nucleon WFs of $s-$ and $p-$shells,
the products of WFs with different permutations do not contribute 
to the integral $\langle s^4 p^{A-4} | s^4 p^{A-4} \rangle$,
which ensures its equality to one.

Using the usual $b$-particle FPCs of Refs.~\cite{Jahn51,Elliott53}, one can decompose the WF of $p$-shell nucleons
giving:
\begin{eqnarray}
 && |s^4 p^{A-4} \alpha LST (x_1,x_2,\ldots,x_{A}) \rangle =
  \left(\begin{array}{c}
    A \\
    4
  \end{array}\right)^{-1/2} \sum (-1)^{\eta_{i_1,\ldots,i_A}} \psi_s(x_{i_1},x_{i_2},x_{i_3},x_{i_4}) \nonumber \\
  && \times \sum \langle p^{A-4} \alpha LST | p^{A-b-4} \alpha_B L_BS_BT_B; p^b \alpha_X {\cal L}S_XT_X \rangle \nonumber \\
  && \times [|p^{A-b-4} \alpha_B L_BS_BT_B (x_{i_5},\ldots,x_{i_{A-b}})\rangle
             \otimes |p^b \alpha_X {\cal L}S_XT_X (x_{i_{A-b+1}},\ldots,x_{i_{A}})\rangle]_{LST}~,  \label{WF_A_split} 
\end{eqnarray}
where $\otimes$ means vector couplings $\bvec{L}=\bvec{L}_B+\bvec{\cal L}$, $\bvec{S}=\bvec{S}_B+\bvec{S}_X$,
and $\bvec{T} = \bvec{T}_B+\bvec{T}_X$.

On the other hand, since the shell model WF of the nucleus $A$ contains the minimum number of the oscillator quanta
compatible with Pauli principle (i.e. $N=A-4$) one can apply the Bethe-Rose-Elliott-Skyrme (BRES) theorem
\cite{Bethe37,Elliott55} and write:
\begin{equation}
  |s^4 p^{A-4} \alpha LST (x_1,x_2,\ldots,x_{A}) \rangle
  = \psi_{00}(\bvec{R}_A) |A \alpha LST (x_1,x_2,\ldots,x_{A}) \rangle~,   \label{WF_A_BRES}
\end{equation}
where $\psi_{00}(\bvec{R}_A)$ is the lowest HO state of the c.m. motion of the nucleus $A$.
By using Eq.(\ref{defFPC}), one can rewrite Eq.(\ref{WF_A_BRES}) in a form with $b$-particle TISM FPCs:
\begin{eqnarray}
  && |s^4 p^{A-4} \alpha LST (x_1,x_2,\ldots,x_{A}) \rangle
  = \psi_{00}(\bvec{R}_A) \sum 
    \langle A \alpha LST | (A-b) \alpha_B L_BS_BT_B; n\Lambda, b \alpha_X  L_XS_XT_X\{{\cal L}\} \rangle \nonumber \\
    && \times
   [|(A-b) \alpha_B L_BS_BT_B (x_1,\ldots,x_{A-b})\rangle \nonumber \\
    &&  \otimes [\psi_{n\Lambda}(\bvec{R}_B - \bvec{R}_X) \otimes |b \alpha_X L_XS_XT_X (x_{A-b+1},\ldots,x_{A}) \rangle]_{\cal L}]_{LST}~, \label{WF_A_BRES_split}
\end{eqnarray}
where we again used short-hand notations $\otimes$ for vector couplings $\bvec{\cal L} = \bvec{\Lambda} + \bvec{L}_X$, $\bvec{L}=\bvec{L}_B+\bvec{\cal L}$,
$\bvec{S}=\bvec{S}_B+\bvec{S}_X$, and $\bvec{T} = \bvec{T}_B+\bvec{T}_X$. 

In a similar way, we can now write the shell model WF of the nucleus $B=A-b$, first, by using the antysymmetrized product, i.e.
\begin{eqnarray}
  && |s^4 p^{A-b-4} \alpha_B L_BS_BT_B (x_1,x_2,\ldots,x_{A-b}) \rangle =
  \left(\begin{array}{c}
    A-b \\
    4
  \end{array}\right)^{-1/2} \sum (-1)^{\eta_{i_1,\ldots,i_{A-b}}} \psi_s(x_{i_1},x_{i_2},x_{i_3},x_{i_4}) \nonumber \\
  && \times  |p^{A-b-4} \alpha_B L_BS_BT_B (x_{i_5},\ldots,x_{i_{A-b}})\rangle~,  \label{WF_B}   
\end{eqnarray}
and, second, by using the BRES theorem, i.e.
\begin{equation}
  |s^4 p^{A-b-4} \alpha_B L_BS_BT_B (x_1,x_2,\ldots,x_{A-b}) \rangle
  = \psi_{00}(\bvec{R}_B) |(A-b) \alpha_B L_BS_BT_B (x_1,x_2,\ldots,x_{A-b}) \rangle~.   \label{WF_B_BRES}
\end{equation}

Let us formally consider the following integral:
\begin{eqnarray}
  I &\equiv& \langle s^4 p^{A-4} \alpha LST (x_1,x_2,\ldots,x_{A}) 
  | [s^4 p^{A-b-4} \alpha_B L_BS_BT_B (x_1,x_2,\ldots,x_{A-b}) \nonumber \\
  && \otimes [\psi_{n\Lambda}(\bvec{R}_X) \otimes b \alpha_X L_XS_XT_X (x_{A-b+1},\ldots,x_{A})]_{\cal L}]_{LST} \rangle~,    \label{Idef}
\end{eqnarray}
where the vector couplings $\bvec{\cal L} = \bvec{\Lambda} + \bvec{L}_X$, $\bvec{L}=\bvec{L}_B+\bvec{\cal L}$,
$\bvec{S}=\bvec{S}_B+\bvec{S}_X$, and $\bvec{T} = \bvec{T}_B+\bvec{T}_X$ are applied.  
Substituting Eqs.(\ref{WF_A_split}),(\ref{WF_B}) into Eq.(\ref{Idef}) and using the orthonormality of the HO shell model basis WFs  we have:
\begin{eqnarray}
  && I =    \left(\begin{array}{c}
    A-b \\
    4
  \end{array}\right)^{1/2}
  \left(\begin{array}{c}
    A \\
    4
  \end{array}\right)^{-1/2}
  \langle p^{A-4} \alpha LST | p^{A-b-4} \alpha_B L_BS_BT_B; p^b \alpha_X {\cal L}S_XT_X \rangle \nonumber \\
  && \times
     \langle p^b \alpha_X {\cal L}S_XT_X (x_{A-b+1},\ldots,x_{A}) 
      | [\psi_{n\Lambda}(\bvec{R}_X) \otimes b \alpha_X L_XS_XT_X (x_{A-b+1},\ldots,x_{A})]_{\cal L} \rangle~.   \label{I_usualFPC}
\end{eqnarray}
One can also express $I$ in a different way, by substituting Eqs.(\ref{WF_A_BRES_split}),(\ref{WF_B_BRES})
into Eq.(\ref{Idef}), which gives:
\begin{eqnarray}
  && I =  \langle A \alpha LST | (A-b) \alpha_B L_BS_BT_B; n\Lambda, b \alpha_X  L_XS_XT_X\{{\cal L}\} \rangle  \nonumber \\
  && \times \langle \psi_{00}(\bvec{R}_A) \psi_{n\Lambda}(\bvec{R}_B - \bvec{R}_X) | \psi_{00}(\bvec{R}_B) \psi_{n\Lambda}(\bvec{R}_X \rangle~. \label{I_tismFPC}
\end{eqnarray}
The last factor in the r.h.s. of Eq.(\ref{I_tismFPC}) is equal to the corresponding
generalized Talmi coefficient (see Eqs.(VI.10),(VI.19) in Ref.~\cite{NS}):
\begin{eqnarray}
  && \langle \psi_{00}(\bvec{R}_A) \psi_{n\Lambda}(\bvec{R}_B - \bvec{R}_X) | \psi_{00}(\bvec{R}_B) \psi_{n\Lambda}(\bvec{R}_X \rangle
  = \langle 00,n\Lambda:\Lambda|\frac{A-b}{b}|00,n\Lambda:\Lambda \rangle  \nonumber \\
  &&  = \left(\frac{A-b}{A}\right)^{n/2} (-1)^n~.   \label{TalmiCoeff}
\end{eqnarray}
By combining Eqs.(\ref{I_usualFPC}),(\ref{I_tismFPC}),(\ref{TalmiCoeff}) one obtains the required relation, i.e. Eq.(\ref{FPC_TISM_vs_usual}).

\end{document}